\documentstyle{mn}
\input{epsf}

\def\la{\mathrel{\mathpalette\fun <}}

\def\fun#1#2{\lower3.6pt\vbox{\baselineskip0pt\lineskip.9pt
\ialign{$\mathsurround=0pt#1\hfill##\hfil$\crcr#2\crcr\sim\crcr}}}

\def\Mpc{{\,h^{-1}\,{\rm Mpc}}}
\def\mpc {h^{-1} {\rm{Mpc}}}
\def\and  {\it {et al.} \rm}
\def\rmd {\rm d}

\def\spose#1{\hbox to 0pt{#1\hss}}
\def\simlt{\mathrel{\spose{\lower 3pt\hbox{$\mathchar"218$}}
     \raise 2.0pt\hbox{$\mathchar"13C$}}}
\def\simgt{\mathrel{\spose{\lower 3pt\hbox{$\mathchar"218$}}
     \raise 2.0pt\hbox{$\mathchar"13E$}}}
\def\beq{\begin{equation}}
\def\eeq{\end{equation}}
\def\bce{\begin{center}}
\def\ece{\end{center}}
\def\bea{\begin{eqnarray}}
\def\eea{\end{eqnarray}}
\def\ben{\begin{enumerate}}
\def\een{\end{enumerate}}

\def\brr{\begin{array}}
\def\err{\end{array}}

\def\calD{{\cal D}}
\def\La{{\cal L}}

\def\bj {b_{\rm J}}
\def\arcsq {{\rm arcsec}^2}

\begin{document}

\title[On the CCD Calibration of Zwicky galaxy magnitudes]
{On the CCD Calibration of Zwicky galaxy magnitudes \\
\& The Properties of Nearby Field Galaxies}

\author[E. Gazta\~{n}aga \& G.B. Dalton]
{E. Gazta\~{n}aga$^{1}$ \vspace{1mm} and G. B. Dalton$^{2}$\\
$^1$Consejo Superior de Investigaciones Cient\'{\i}ficas (CSIC), 
Institut d'Estudis Espacials de Catalunya (IEEC), \\ 
Edf. Nexus-104 - c/ Gran Capitan 2-4, 08034 Barcelona, Spain.\\
$^2$Astrophysics, University of Oxford, Keble Road, Oxford, OX1 3RH, UK.}

\maketitle 
\def\magsq {\;{\rm mag\;arcsec}^{-2}}
\def\mag {\;{\rm mag}}
\def\mpc {h^{-1} {\rm Mpc}}
\def\impc {h {\rm Mpc}^{-1}}
\def\and  {{\it {et al.} }}
\def\rmd {{\rm d}}

\begin{abstract}

We present CCD photometry for galaxies
around $204$ bright ($m_Z < 15.5$) Zwicky galaxies 
in the equatorial extension of the APM Galaxy Survey,
sampling and area over 400 square degrees,
which extends 6 hours in right ascension. 
We  fit a  best linear relation between
the Zwicky magnitude system, $m_Z$, and the CCD photometry, $B_{CCD}$, 
by doing a likehood analysis that corrects 
for Malmquist bias. This fit yields a mean scale error
in Zwicky of 0.38 mag per magnitude: ie
$\Delta m_Z \simeq ~(0.62 \pm 0.05) ~\Delta B_{CCD}$
and a mean zero point of $<B_{CCD}-m_Z>= -0.35 \pm 0.15$ mag. The scatter
around this fit is about $0.4$ mag. 
Correcting the Zwicky magnitude system with the best fit model
results in a 60\% lower normalization  and $0.35\mag$  brighter $M_*$ 
in the luminosity function. This brings the CfA2 luminosity
function closer to the other low redshift estimations 
(eg Stromlo-APM or LCRS).
We find a significant positive angular correlation of
magnitudes and position in the sky at scales smaller than about 5 armin,
which corresponds to a mean separation of $120h^{-1}\;{\rm Kpc}$. 
We also present colours, sizes and ellipticities for galaxies
in our fields which  provides a good local reference for 
the studies of galaxy evolution. 

\end{abstract}


\section{Introduction}

Some important aspects of galaxy evolution can only be understood by
studying the statistical properties of nearby field galaxies, in particular
its luminosity function (LF). As well as providing vital information
for galaxy evolution studies, an accurate knowledge of the present day
LF is needed to normalize the number counts of galaxies at fainter
magnitudes, and to understand the clustering and large scale
structure of the galaxy distribution.
Moreover, the colour distribution of the nearby galaxies
provides a basic reference to determine star formation rates
in galaxies.

Much recent work has been directed towards studying the evolution of
the LF using samples of faint galaxies at high-redshift, but a
consideration of the ensemble of available estimates of the LF at low
redshifts suggests that there are still large inconsistencies which
must be reconciled before reliable conclusions can be drawn about the
implications of deep surveys.

\begin{table*}
\begin{tabular}{lcrlccc}\\ 
Survey & Band & $N_{gal}$ & $\bar{z}$ & $M_*$ & $\alpha$ & $\phi_*$ \\ \hline
       &          &           &           &       &          & $h^3\;{\rm Mpc}^{-3}$ \\ \hline
CfA2 & $m_Z$    &  9063 & 0.025 & $-18.8\pm0.3$ & $-1.0\pm0.2$ & $0.040\pm0.01$ \\
SSRS2 & $B(0)$  &  2919 & 0.025 & $-19.5\pm0.08$ & $-1.2\pm0.07$ & $0.015\pm0.003$ \\
SAPM &  $\bj$   &  1658 & 0.050 & $-19.5\pm0.13$ & $-1.0\pm0.15$ & $0.014\pm0.002$ \\
ESP & $\bj$     &  3342 & 0.1   & $-19.6\pm0.07$ & $-1.2\pm0.06$ & $0.020\pm0.004$ \\
2dF & $\bj$     &  5869 & 0.14  & $-19.7\pm0.06$ & $-1.3\pm0.05$ & $0.017\pm0.002$ \\
LCRS & $r$      & 18678 & 0.1   & $-20.3\pm0.02$ & $-0.7\pm0.05$ & $0.019\pm0.001$ \\
CS &  $R_{KC}$  &  1762 & 0.06  & $-20.7\pm0.2$ & $-1.2\pm0.2$ & $0.025\pm0.006$ \\ \hline
\end{tabular}
\caption[dum]{\label{tb:lfpars} Luminosity Function parameters derived
from currently available surveys. }
\end{table*}

It is common practice to fit the LF to the so called
Schechter (1976) form:
\beq
\phi(L) = \phi_* ~ \left({L\over{L^*}} \right)^\alpha ~ {\rm exp}
\left(-{L\over{L^*}}\right),
\label{schechter}
\eeq
were luminosity is related with magnitude in the usual way,
${L\over{L^*}}=10^{0.4(M_*-M)}$. Determinations of the low redshift
$B$-band LF are available from the Stromlo-APM Survey (SAPM, Loveday et al.,
1992), the Southern Sky Redshift Survey (Da Costa et al., 1994), the
CfA2 redshift survey (Marzke et al., 1994), and more recently from the
ESO Slice Project (Zucca et al., 1997) and the 2dF Galaxy Redshift
Survey (Folkes et al., 1999). Determinations in the $R$-band are
available from the Las Campanas Redshift Survey (LCRS, Lin et al., 1994) and
from the Century Survey (CS, Geller et al., 1997). The best determinations
of the Schechter function parameters from each of these surveys are
listed in Table~\ref{tb:lfpars}. The SSRS2 is based on the
ESO-Uppsala photometry of Lauberts \& Valentijn (1989), but
transformed to the $B(0)$ system of Huchra (1976). Da Costa et
al. (1994) ascribe the difference in the LFs of the two surveys to the
large colour-term used by Lauberts \& Valentijn to relate the two
systems, and so we therefore use $m_Z$ for Zwicky magnitudes and
$B(0)$ for SSRS2 magnitudes. Da Costa et al. (1994) quote the $rms$
difference between $B(0)$ and $\bj$ as $\simlt 0.2 \mag$. We quote the
results for the red LFs of the Las Campanas and Century surveys here
for completeness, and do not concern ourselves with the details of the
transformations between red and blue passbands. We note however that
Geller et al., 1997 find the LF obtained from the CS to be in
excellent agreement with a prediction based on colour transfmorations
applied to the SSRS2 LF.

\begin{figure}
\centerline{\epsfxsize=7.truecm \epsfbox{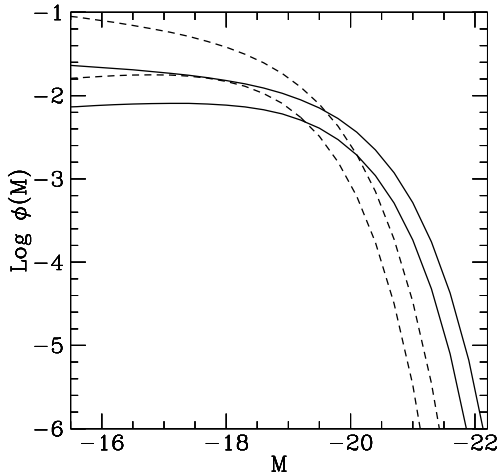}}
\caption{Luminosity function estimations in the SAPM and
the CfA2 as a function of absolute magnitudes $M$ in each magnitude
system ($b_J$ or $m_Z$). The continuous lines enclose the 2-sigma region in
the SAPM estimation whereas the dashed lines enclose the
2-sigma region in the CfA2 estimate.}
\label{lf}
\end{figure}

Allowing for this adjustment between blue and red magnitudes, the
final six rows of Table~\ref{tb:lfpars} are all in reasonable
agreement, with only the CfA2 result showing a large deviation,
particularly in the value of $\phi_*$. The difference in the LFs
deduced from the CfA2 and SAPM surveys is illustrated in
Figure~\ref{lf}.

At face value, the CfA2 LF seems to have less intrinsically bright
galaxies (at least 10 times less M=-21 galaxies). Of course, this
depends on the transformation between Zwicky $m_Z$, APM $b_j$ or LCRS
$R$ magnitudes.  Efstathiou {\it et al.} (1988) adopted the relation
$m_Z=B +0.29$ for the CfA1 survey (Huchra et al. 1983). This
transformation would move the CfA2 LF in the correct direction to
reconcile it with other measurements, although Lin et al. (1994) point
out that a shift of $0.7\mag$ in $M$ is required to reconcile $M_*$
with the other surveys. However, a simple shift in the magnitude
zero-point would not correct for the apparent discrepancy in the
normalisation, which would suggest that a more subtle effect is
present in the CfA2 data.

In this paper we present the results of a photometric survey of bright
galaxies in the overlap region of CfA2 and the equatorial extension of
the APM Galaxy Survey (Maddox et al. 1990a,b; Maddox et al. 1991),
which we will use to investigate this apparent discrepancy. In
Section~2 we begin by comparing the APM photometry with the Zwicky
measurements. We describe our CCD observations in Section~3, and
compare our observations with Zwicky's photometry in Section~4. In
Section~5 we give some other properties for the galaxies in our
fields and dicuss the implications for the LF.  
In Section~6 we compare our results with the findings of
other authors and discuss the possible implications for other results.
No direct CCD calibration has yet been published for this part of the
APM Galaxy Survey. We will present a detailed comparison of our CCD
photometry with the APM data down to the survey completeness limit in
a separate paper.

\section{A Comparison of Zwicky and APM magnitudes}
\label{ssapmcomp}

From the information used to calibrate the APM survey we would expect
$B\simeq \bj^{\rm APM}+0.2$, given the mean $(B-V)\sim 0.7$ and the
colour equation $\bj = B - 0.28(B-V)$ (Blair \& Gilmore 1982; Maddox
et al. 1990b; although this shift could increase by as much as
$0.07\mag$ according to the findings of Metcalfe, Fong \&
Shanks, 1995). Adopting the relation between $m_Z$ and $B$ used by
Efstathiou, Ellis \& Peterson (1988), this transformation would imply
\beq
\label{eq-req}
m_Z \simeq \bj + 0.5, 
\eeq
which is close to the shift in $M_*$ adopted by
Lin et al. (1996), and, if taken as they stand, would suggest that the
results of the two surveys might be consistent, at least in $M_*$.

\begin{figure}
\centerline{\epsfxsize=7.truecm \epsfbox{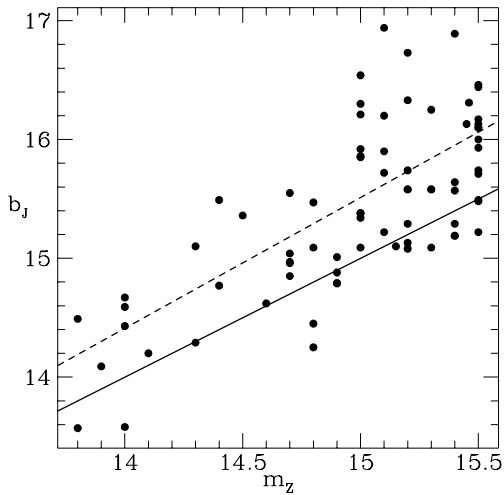}}
\caption{Comparison of the apparent magnitudes for a sample of $\sim
100$ galaxies in the overlap region of the SAPM and the CfA2.
Magnitudes in the SAPM catalogue correspond to APM $b_J$
estimations while magnitudes in the CfA2 are Zwicky magnitudes,
$m_Z$.  The continuous line corresponds to $b_J=m_z$ and the dashed
line is the best linear fit, $m_Z= b_J-0.6 \pm 0.5$, showing both an
offset and a large scatter. We would expect to find at least
$m_Z=b_J+0.5$}
\label{mzbj}
\end{figure}

We investigated the usefulness of this relation by considering the
subset of $\sim 100$ galaxies found in the part of the APM Galaxy
Survey equatorial extension which overlaps with the Southernmost
region of CfA2 (The S+3 sample of Marzke et al., 1994). This sample is
drawn entirely from volume V of the Zwicky catalogue. We have used
the publically available CfA1 catalogue as our
source for Zwicky galaxies, with magnitudes corrected as described 
in CfA1. The CfA1 includes redshifts
only for galaxies with $m_Z < 14.5$, but has magnitudes and positions
for $m_Z < 15.5$. Marzke et al
find this sample to be representative of the whole of the Southern
part of CfA2 (their Figure~3). We inspected this sample of galaxies on
the film copies of the UKST IIIa-J Sky Survey plates, and examined the
image maps as reconstructed from the APM survey data. We removed from
our sample all those galaxies which had either been broken up into
multiple image components or which were composed of multiple
components which had been merged into a single object by the APM
software. The distribution of these objects in the $m_Z,\bj$ plane is
shown in Figure~\ref{mzbj}. A simple least squares fit to these data with the
slope constrained to be unity gives a zero-point shift of 
\beq
\label{eq-obs}
m_Z = \bj - 0.6\pm0.5,
\eeq
which is more than a magnitude in the opposite sense to that implied
by equation \ref{eq-req}. As will be noted below a detailed
calibration requires a correction for Malmquist bias.

The APM Survey is internally calibrated by matching images in plate
overlaps, with the overall zero-point fixed by a number of CCD frames
distributed over the survey region (Maddox et al., 1990b). At bright
magnitudes ($\bj\simlt17$) the calibrations are less well determined
due to a combination of variations in galaxy surface brightness
profiles and the smaller number of bright galaxies found in the plate
overlap regions. The equatorial survey extension has been calibrated
by matching to the original survey using plate overlaps and the
adopted zero-point for the whole survey taken from the original CCD
photometry.  Maddox et al. (1990b) also used their CCD frames to
perform an internal consistency check on the quality of the
plate-plate matching technique, and found the residual zero point
errors on individual plates to be $\simlt 0.04\mag$. The individual
uncertainties in galaxy magnitudes in the APM system are therefore
expected to be much smaller than effect shown in Figure~\ref{mzbj}. It
is possible that an improvement in the quality of the plate material
used for the more recent plates of the equatorial extension coupled
with changes in the plate copying process could result in a small
change in the saturation correction required for bright galaxies, and
that this could produce an effect in this direction (Maddox, private
communication). However, account was taken of such effects in the
construction of the equatorial extension, and any residual effect is
likely to be much smaller than the discrepancy shown here.

The transformation of Zwicky magnitudes to $B$ deduced by
Efstathiou, Ellis \& Peterson (1988) are based on a comparison of the
CfA1 survey data with the Durham--AAT redshift survey using 139
galaxies (Peterson et al. 1986). These authors find that the different
volumes of the Zwicky catalogue have large variations in the
zero-points, although volume V is found to be representative of the
calibration as estimated from all volumes.  These data are limited to
$m_Z < 14.5$, and the luminosity function parameters inferred for the
CfA1 are consistent with the parameters listed for the other surveys
in Table~1. It is not possible to draw a direct comparison of the APM
data with Zwicky data brighter than $m_Z \sim 14$, due to heavy
saturation of galaxies this bright in the APM data. The most plausible
inference from the comparison described above would therefore be a
difference in the magnitude {\it scales} of the two surveys for $\bj
\simgt 14$, with the calibration of Efstathiou, Ellis \& Peterson
(1988) being appropriate for the Zwicky system at brighter magnitudes.

\section{CCD Data}

\subsection{Observations}
\label{ssccdobs}

We used the galaxies from the sample discussed in the previous section
as the basis for a CCD survey to investigate the above discrepancy by
providing an independent calibration for both surveys.  This overlap
region is essentially defined as $21^{\rm h} 50 < \alpha < 3^{\rm h}
40$ and $-0.25^\circ < \delta < 0.25^\circ$.  We obtained images with
the 2.5m Isaac Newton Telescope(INT) and 1.0m Jacobus Kapteyn
Telescope(JKT) in October 1994. The decision to use two telescopes was
motivated by the number of close groups of Zwicky galaxies in the
region, so that we use the $10^\prime$ field of view available at the
INT to image cluster fields containing groups of bright galaxies and
the $6^\prime$ field of view of the JKT to image individual
galaxies. We used identical Tektronix $1024\times 1024$ detectors
(TEK3 and TEK4) and Harris $B$ and $R$ filters on the two telescopes.  We
obtained observations of 58 fields in two nights at the INT and 73
fields in three nights at the JKT with exposure times of 360s and
600s, respectively. We observed a number of photometric standards from
the list of Landolt (1992), at hourly intervals throughout each night,
as well as the field of the Active Galaxy AKN120 which contains a
number of photometric standards (Hamuy \& Maza, 1989).

The data were reduced using standard techniques as implemented in the
IRAF {\tt ccdred} packages, with the exception that we used a modified
version of the overscan correction routine to compensate for a
saturation of the preamplifier used with TEK4 on the JKT. This effect
manifested itself as a sudden drop in the overscan level following
readout of particularly bright stellar objects in the field, and
subsequent exponential recovery to the normal level as the next $\sim
100$ rows were read out. We found that this problem could be corrected
by fitting the overscan regions on either side of the drop for those
fields where the effect occurred.

Extinction coefficients were derived each night, giving values in the
range $0.10 \le k_B \le 0.12$. We determined zero-points for the two
combinations of telescope and detector to be 
\begin{equation}
B_{0,INT}=24.16 \pm 0.02,
\end{equation}
and
\begin{equation}
B_{0,JKT}=21.77 \pm 0.02. 
\end{equation}
We also obtained $R$-band images of our standards and target fields,
with zero points determined to be $R_{0,INT}=24.33 \pm 0.1$ and
$R_{0,JKT}=22.11 \pm 0.02$, but we were unable to determine any
significant colour term from our standard stars, consistent with
previous experience with this combination of CCD and filters which are
designed to be very close to the Johnson--Cousins system.

\subsection{Image Detection and Photometry}
\label{sspisa}

\begin{figure}
\centering
\centerline{\epsfxsize=8.truecm 
\epsfbox{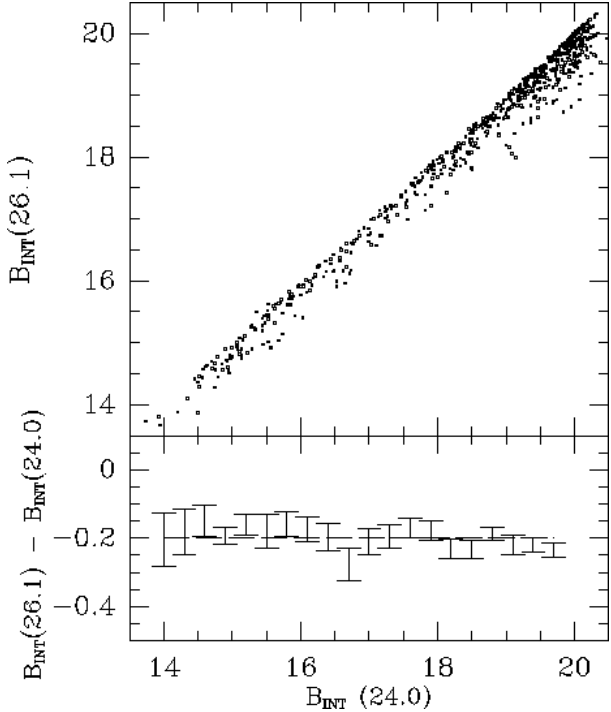}}
\caption[junk]{Comparison of total magnitudes $B_{INT}$ for different
isophotes: $I=26.1\magsq$ to $I=24.0\magsq$. 
}
\label{ccdbt}
\end{figure}

We used the STARLINK {\tt PISA} image detection and photometry
software for our photometric analysis.  This is essentially the same
as the image detection software used in the construction of the APM
Galaxy Survey (Irwin, 1986). The basic input parameters for the
detection are the threshold intensity per pixel or surface brightness,
$I_{s}$, and the minimum size of the object to be detected,
$A_{s}$. Amongst other parameters PISA returns total area $A_T$, 
the ellipticities and the
isophotal magnitude $B_I$ or the corresponding total magnitude $B$
resulting from a curve of growth analysis for each detected object
after removal of any overlapping objects.

Given that our CCD survey was designed to provide a calibration of
both the APM and Zwicky data, we were necessarily interested in a wide
range of magnitudes ($14 \le B \le 20$) and image sizes. For this
range of objects there was no unique combination of $I_{s}$ and
$A_{s}$ that could deblend the faint objects without breaking the
bright ones.  In order to automate this process as much as possible
for the whole range of magnitudes, we ran {\tt PISA} several times
with different combinations of $I_{s}/A_{s}$. These are listed in
Table~\ref{tb:sizes}. 

\begin{table}
\begin{center}
\begin{tabular}{lrr} \\
 & $I_s$ & $A_s$ \\
 & $\magsq$ & $\arcsq$ \\ 
INT1 &  26.1 & 96 \\
INT2 &  25.7 & 34 \\
INT3 &  25.4 & 34 \\
INT4 &  25.0 & 17 \\
INT5 &  24.6 & 10 \\
INT6 &  24.4 & 10 \\ \hline
JKT1 &  25.0 & 12 \\
JKT2 &  24.5 & 12 \\
JKT3 &  24.1 &  6 \\
JKT4 &  23.7 &  6 \\
JKT5 &  22.9 &  3 \\ \hline
\end{tabular}
\caption[dum]{\label{tb:sizes} Input parameter sets used for image
analysis.}
\end{center}
\end{table}


We chose a larger (smaller) area for the fainter (brighter) isophotes
so as to select similar objects in all runs.  Objects in the final
catalogue were selected from the INT4 and JKT4 runs. The information
from different isophotes was then used to perform an automatic
rejection of broken or contaminated images.  After rejection, the
total magnitude and size were the largest remaining estimates of $B$
and $A_T$, which typically correspond to the faintest isophote
left. The error in $B$ was taken to be the variance in the different
estimates for total magnitudes. Objects with rejected isophotes were
automatically labeled and checked visually.  We refer to total
magnitudes estimated in this way as $B_{INT}$ and $B_{JKT}$ for the
INT and JKT sets, respectively, or $B_{CCD}$ in general. These
effectively correspond to total magnitudes detected at isophotes
$26.1\magsq$ and $25.0\magsq$, respectively, unless stated otherwise.


To check if our isophotes are low enough we have also computed
total magnitudes determined by PISA using higher
isophotes.
We find a small zero-point shift of the total $B_{INT}$ scale as
a function of the isophote $I$, but for a given $I$ this shift is not
a function of $B$ over the wide magnitude range considered
here. This is illustrated in Figure \ref{ccdbt} which shows a change
of $\approx 0.2\mag$ in the mean $B_{INT}(I)$ when we change the
isophote from $I=26.1\magsq$ to $I=24.0\magsq$.  Changing the isophote
from $I=26.1\magsq$ to $I=25\magsq$ introduces a change of only
$0.1\mag$.  Changing the detection isophote from from $I=26.1\magsq$
to $I=24.4\magsq$ also introduces a change in $B_{INT}$ of about
$0.2\mag$. We therefore conclude that the residuals associated with
our final choice of detection isophote are small compared to the
shifts illustrated in Figure~\ref{mzbj}.

This analysis also implies that there should be a mean zero-point
shift in the JKT data relative to the INT data of $0.1\mag$ due to
difference in the isophotes used. We therefore apply this shift to all
objects observed with the JKT to transform these data to our
$B_{26.1}$ system.


%

\begin{figure}
\centering
\centerline{\epsfxsize=8.truecm 
\epsfbox{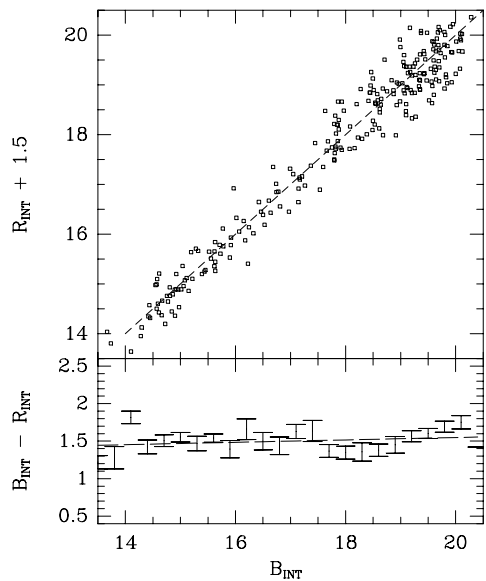}}
\caption[junk]{The top panel shows the red magnitudes $R_{CCD}$
for a subsample of INT galaxies as a function of the blue $B_{CCD}$ ones.
The bottom panel shows
the colour $B_{CCD}$-$R_{CCD}$ evolution as a function $B_{CCD}$.}
\label{intrb}
\end{figure}

Finally, we investigated the possibility that there could be a small effect
due to the change in mean redshift of the samples at fainter magnitudes
by searching for a change in the mean galaxy colour.
Figure~\ref{intrb} shows ($B_{INT}$-$R_{INT}$) as a function of
$B_{INT}$. There is only a small colour evolution within the errors.
A linear fit to the binned data in Figure~\ref{intrb} yields $B-R= 
0.016~B + 1.23$ with a mean $B - R = 1.50$ (the mean weighted by
each galaxies is  $B - R = 1.61$ as is dominated by the faint objects).

\section{Comparison with the Zwicky Catalogue}
\label{ssmatch}

\begin{figure}
\centering
\centerline{\epsfxsize=8.truecm 
\epsfbox{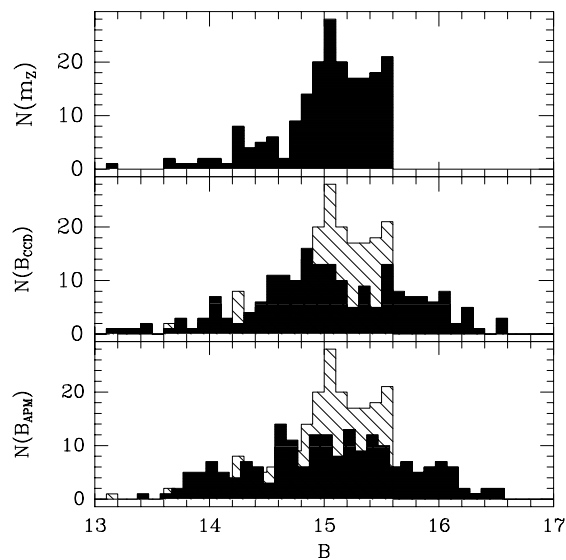}}
\caption[junk]{Histograms comparing the numbers of galaxies as a
function of apparent magnitude $B$. Each panel shows in black, from
top to bottom, the Zwicky, APM and CCD magnitudes for the same
objects. In the lower two panels we also show the Zwicky histogram as a
background for comparison.}
\label{zwhist}
\end{figure}

In Figure \ref{zwhist} we show the number counts histograms for the
204 Zwicky galaxies in our sample to the different magnitude systems:
Zwicky (top), APM (middle), and CCD (bottom). Comparison of the APM
and CCD data shown here suggests that the effect shown in
Figure~\ref{mzbj} is unlikely to be an artefact of saturation effects
in the equatorial APM Survey data at bright magnitudes (see
Section~\ref{ssapmcomp}).

\begin{figure}
\centering
\centerline{\epsfxsize=8.truecm 
\epsfbox{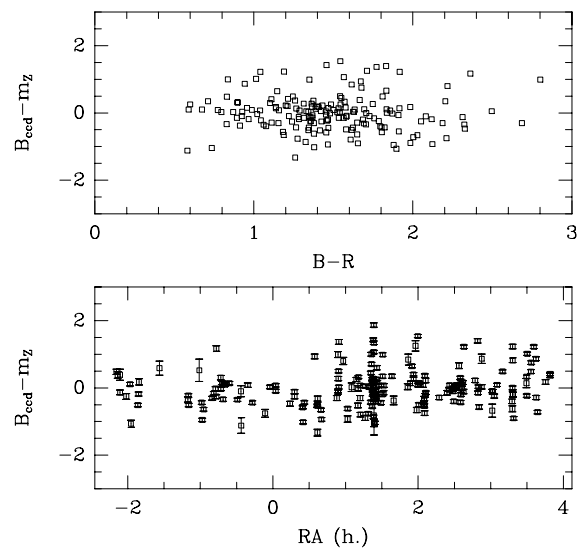}}
\caption[junk]{$B_{CCD}-m_Z$ as a function of
CCD colour $B-R$ (top panel) and right ascention (bottom).}
\label{zwccddra}
\end{figure}

We searched for possible correlations between the magnitude
differences, $B_{CCD}-m_Z$, and other properties of our sample.  The
top panel of Figure~\ref{zwccddra} shows the $B_{CCD}-m_Z$ difference
as a function of CCD colour $B-R$.  The colour of Zwicky galaxies is
similar to the mean colour in Figure \ref{intrb}, $B-R \simeq 1.5$ and
there is no apparent correlation with $B_{CCD}-m_Z$ within the
scatter. This is a useful check, as any large differences that were
due to processing errors might be expected to show up as objects of
extreme colour. The bottom panel of Figure~\ref{zwccddra} shows the
scatter in $B_{CCD}-m_Z$ as a function of right ascension. Objects
were observed in order of increasing RA on each night of our observing
run, and so we would expect temporal drifts to show up as a
correlation with RA. Again, there is no evidence of any trend. We also
investigated possible variations with Galactic latitude (not shown
here), and again found no evidence for trends in our data.

\begin{figure}
\centering
\centerline{\epsfxsize=8.truecm 
\epsfbox{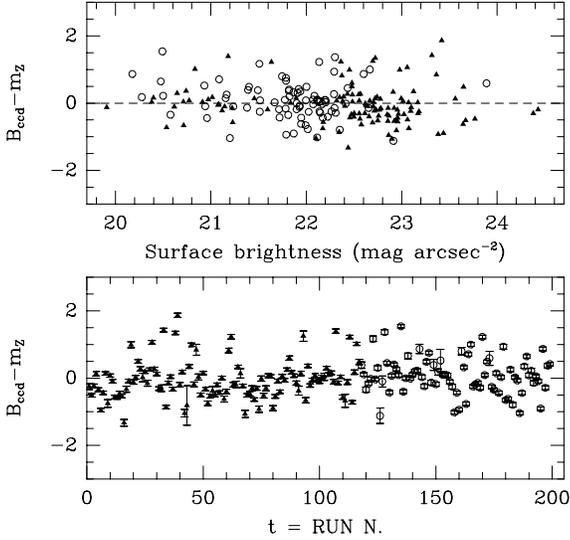}}
\caption[junk]{$B_{CCD}-m_Z$ as a function of
surface brightness (top panel) and run number or time
of observation (bottom).}
\label{zwccddrun}
\end{figure}

The top panel of Figure \ref{zwccddrun} shows the $B-m_Z$ error as a
function of the mean surface brightness for all Zwicky galaxies.
Surface brightness is defined here as the ratio of isophotal
magnitudes to the area above a threshold of $25.0\magsq$ in the INT
(triangles) and with a threshold of $24.1\magsq$ in the JKT
(circles). The difference in the threshold explains the systematic
shift in the bulk value of mean high surface brightness of the JKT
images (as we are closer to the galaxy core). As can be seen in the
Figure, there is no apparent variation of magnitude error with surface
brightness.

The bottom panel of Figure \ref{zwccddrun} shows the $B-m_Z$ error as
a function of the sequential run number. This is close to
observational time for the INT (triangles) or JKT (circles) when
considered separately (in practice, some of the INT and JKT
observations were made on the same night).  There is no apparent
variation of magnitude error with run number.

\begin{figure}
\centering
\centerline{\epsfxsize=9.truecm 
\epsfbox{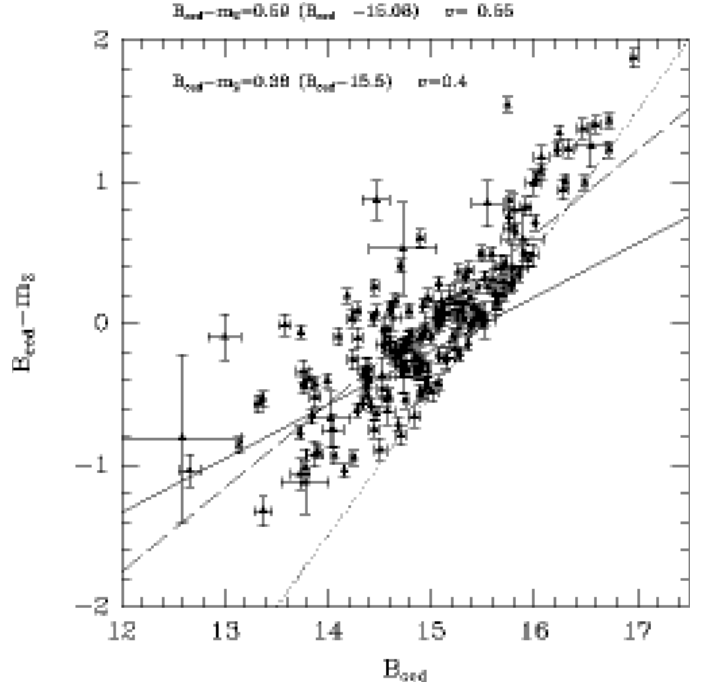}}
\caption[junk]{The magnitude error $B_{CCD}-m_Z$ as a function of
$B_{CCD}$. The dotted
line shows the $m_Z=15.5$ magnitude limit. The dashed line corresponds
to a direct least square fit to the data. The continuous line shows
the fit corrected for Malmquist bias (see text).}
\label{zwccdd}
\end{figure}

Figure \ref{zwccdd} is the main result of this paper.  It shows the
Zwicky magnitude error $B_{CCD}-m_Z$ as a function of $B_{CCD}$, for
all Zwicky galaxies with $B_{CCD}<17.5$.  For the Zwicky magnitudes we
have used a constant error: $\Delta m_Z =0.05\mag$ (as Zwicky
quoted magnitudes with a precision of $0.1\mag$).  For the CCD data we used
the error described in Section~\ref{sspisa}.  A very similar trend is
found for the separate INT (closed circles) and JKT images (open squares).

The dashed line in Figure~\ref{zwccdd} shows a direct least square
fit to the data:
\beq 
B_{CCD}-m_Z \simeq  0.6~ (B_{CCD}-15.1),
\label{lsqf}
\eeq which has a scatter of $0.55\mag$.  

This fit is the result of the interplay between the scatter in the two
magnitude systems and the Zwicky survey limit (shown as dotted line in
Figure~\ref{zwccdd}).  As a result of this scatter, objects with faint
$B_{CCD}$ and bright $m_Z$ can be included in the survey, but there is
a deficit of objects with faint $m_Z$ and bright $B_{CCD}$. i.e. the
fit suffers from a Malmquist type of bias. However, for a linear
relation, it is apparent that Malmquist bias is insufficient to
account for all of the effect shown in Figure~\ref{zwccdd}.  We next
develop a scheme to correct this fit for the effects of Malmquist bias.

\subsection{Malmquist bias correction}

Different magnitude system are subject to different systematic
errors, and even in the best of the situations intrinsic differences in
the morphology, environment and spectrum of the galaxies tend to introduce 
stochastic fluctuations in any magnitude system. We want to
find a best fit linear relation between the Zwicky, ${m_Z}$, and 
CCD, $B_{CCD}$, system:
\beq
m_Z= \lambda B_{CCD} + Z,
\label{linear}
\eeq 
where $\lambda$ will account for any scale difference and $Z$ is
a zero point shift.  In general, both $m_Z$ and $B_{CCD}$ are
stochastic variables and equation~\ref{linear} is just a mean relation.
As is common practice we will assume that there is Gaussian scatter
around the above mean relation (due to the accumulation of multiple
uncorrelated factors). That is, given a measured magnitude $m_Z$,
the error is given by:

\beq
P(\Delta)={1\over{N}} \,
\exp\left[{-{{\Delta^2}\over{2 \sigma^2}}}\right],
\label{gauss}
\eeq
where $\Delta \equiv m_Z-\overline{m_Z}$ 
is a stochastic variate, $\overline{m_Z}$
is some mean best fit value in the linear relation of equation~\ref{linear},
$\sigma$ is the $rms$ error and $N$ is a normalization factor.
For a sample that is not magnitude limited we have $N=\sqrt{2\pi} \sigma$. 
For a magnitude limited sample, where $m_Z<m_Z^{lim}$,
the probability is the same but has a different normalization:
\beq
N = \int_{-\infty}^{m_Z^{lim}-\overline{m_Z}} \,  dm_Z \,
\exp\left[{-{{\Delta^2}\over{2 \sigma^2}}}\right],
\eeq

because not all magnitudes are possible, so that $\Delta \equiv
m_Z-\overline{m_Z}< m_Z^{lim}-\overline{m_Z}$.  Thus, we can write
$P(m_Z)$ in terms of the complementary error function: 
\beq 
P(\Delta)={2\, \exp\left[{-{{\Delta^2}\over{2 \sigma^2}}}\right]
\over{\sqrt{2\pi}\,\sigma\, {\rm
erfc}\left[{\overline{m_Z}-m_Z^{lim}\over{\sqrt{2} \sigma}}\right]}} \,
\label{pc}
\eeq
so that for $m_Z^{lim}=\infty$ we recover the standard Gaussian result.

We are now able to perform a likelihood analysis to find the best fit
values of $\lambda$ and $Z$ in the linear relation of
equation~\ref{linear}.  We define a likelihood as:

\beq
\La = \prod_i P(\Delta_i)
\eeq
where $i$ runs over all galaxies in the survey, and:
\beq 
\Delta_i = m_Z^i \, -(\lambda \,  B_{CCD}^i \, + \,Z)
\eeq
with $m_Z^i$ and $B_{CCD}^i$ the measured Zwicky and CCD magnitudes for
galaxy $i$. In analogy with the standard $\chi^2$ test we define
a Malmquist bias ``corrected chi-square'':
\beq
\chi^2_{Malm} \equiv \sum_i \left[ {{\Delta^2_i}\over{2 \sigma^2}} + 2
\, {\rm log}({\rm erfc}{\lambda B_{CCD}^i+Z-m_Z^{lim}\over{\sqrt{2} \,
\sigma}}) \right]
\eeq
Note that the measured magnitude errors, $\sigma_i$, are added in 
quadrature to the stochastic error in the linear relation, $\sigma$, which 
is one of the parameters we want to determine with this fit.
The best fit values that we find on maximising the likelihood, using
the data in Figure \ref{zwccdd}, are:
\bea
\lambda &=& 0.62 \pm 0.05 \\ \nonumber
Z &=& 5.9 \pm 0.7 \\
\sigma &=& 0.40 \pm 0.05, \nonumber
\eea
the errors correspond to approximate 99\% confidence levels. 
The values in each uncertainty range are strongly correlated:
Lower values of $\lambda$ (eg larger scale errors) are (linearly)
correlated with larger values of $Z$, and both are obtained 
for the smaller $\sigma$. The inverse relation gives:
\beq
B_{CCD} \simeq 1.61 m_Z -9.5.
\label{bfit}
\eeq
Figure \ref{zwccdd} shows as a  continuous line the best fit model
for the corresponding best fit magnitude difference:
\beq
B_{CCD}-m_Z = (0.38 \pm 0.02)  (B_{CCD}-15.53) \pm 0.4  
\eeq

\begin{figure}
\centering
\centerline{\epsfxsize=7.truecm 
\epsfbox{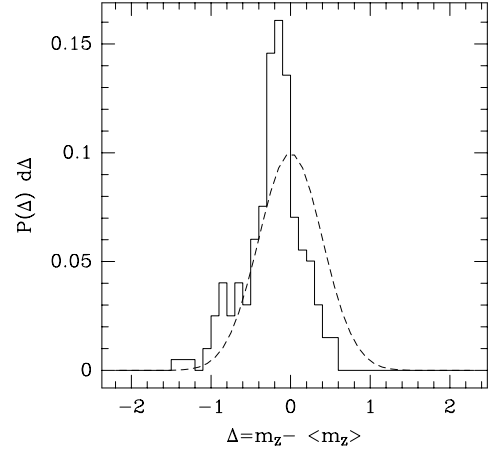}}
\caption[junk]{The distribution of magnitude errors from the mean
linear fit compared to a Gaussian with dispersion $\sigma=0.4$. The
Gaussian does not fit the distribution because of the Malmquist bias,
which produces a deficit of faint objects.}
\label{pdzwc}
\end{figure}

Figure \ref{pdzwc} shows the residuals of this fit. For comparison we
display a Gaussian with same dispersion, $\sigma=0.4$.  As mentioned
above the Gaussian does not provide a good fit because of the Malmquist
bias, which produces a deficit of faint objects.  It is not
possible to show in this Figure a comparison with the Malmquist bias
corrected version for the error probability of equation~\ref{pc}, because
this depends not only on the differences, $\Delta$, but also on the
measured value, $B_{CCD}^i$.

Thus, after correcting for the effects of Malmquist bias, the above
analysis indicates both a zero-point shift and a change of magnitude
scale. The zero point different $Z_0$ could be obtained by taking the mean
of the magnitude differences over $m_Z$ magnitudes (which are the ones
that define the survey limit): 
\beq 
Z_0 \equiv <B_{CCD}-m_Z>= -0.35 \pm 0.15.
\label{zero}
\eeq
which is in good agreement with Efstathiou {\it et al.} (1988).
The scaling relation between the two magnitude systems is then:
\beq
\lambda \equiv {\Delta m_Z\over{\Delta B_{CCD}}} 
~\simeq ~ 0.62 ~\pm~ 0.05 .
\label{scalerror}
\eeq

\begin{figure}
\centering
\centerline{\epsfxsize=9.truecm 
\epsfbox{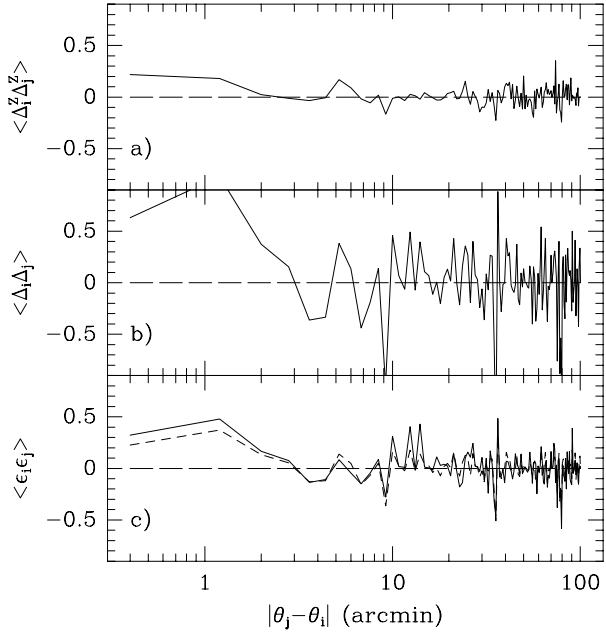}}
\caption[junk]{Mean correlations in magnitude
differences $<\Delta(\theta_i)\Delta(\theta_j)>$ 
for pairs of galaxies separated by
distance $|\theta_j-\theta_i|$, in arcmin shown for Zwicky magnitudes
(top panel), CCD magnitudes (middle panel), and for the difference
$\epsilon$ between Zwicky and CCD magnitudes (lower panel). The
short-dashed line shows the prediction using the scale error model
$\epsilon=0.6 \Delta B$.  The long-dashed line shows zero correlation
for reference.}
\label{nrzw}
\end{figure}

We can also test the above model for the scale error by estimating the
magnitude correlations as a function of projected separation.
Figure~\ref{nrzw} shows the mean correlation
$<\Delta(\theta_i)\Delta(\theta_j)>$ as a function of the pair
separation $|\theta_j-\theta_i|$, in arcmin.  The uppermost panel
shows (as continuous lines) the autocorrelation for magnitude
differences in the Zwicky system, $\Delta^Z_i=m_Z^i -\overline{m_Z}$,
where $m_Z^i$ is the Zwicky magnitude for the $i$th galaxy and
$\overline{m_Z}=<m_Z>$ is the mean Zwicky magnitude in the Survey. The
middle panel shows the corresponding autocorrelation for the CCD
system: $\Delta_i=B_{CCD}^i -\overline{B}_{CCD}$.  The lower panel
shows the autocorrelation of the magnitude errors: $\epsilon_i =
B_{CCD}^i -m_Z^i$.  The long-dashed in each case show the zero values
for reference. The short-dashed line in the lower panel shows the
prediction based on applying equation~\ref{scalerror} to the data,
which is extremely close to the observed result.

We can see in Figure \ref{nrzw} that there is a significant angular
correlation between nearby magnitudes and positions
 with $\theta \la 5^\prime$.  This
correlation is followed by the errors, indicating that the above model
is valid. The Zwicky system shows smaller correlations and smaller
variance than the CCD system. This can also be understood in the
context of the model, as in the Zwicky system the ``effective''
magnitude scale is about a factor of two smaller (equation~\ref{scalerror}).

Note that at the typical depth of the survey, 
$\calD \simeq 80\Mpc$, the above
magnitude correlations are only significant at physical scales smaller
than $\la 100\,h^{-1}\,{\rm Kpc}$.  
This correlation has little effect on typical
galaxy clustering scales, $r_0 \simeq 5 \Mpc$, but might be
relevant for the inversion of angular clustering on
smaller scales. For example, Szapudi \& Gazta\~naga (1998) find
that on scales $\theta \la 0.1^\circ$ there is 
a significant disagreement between the APM and the EDSGC
that is attributed to differences in
the construction of the surveys, most likely the dissimilar 
deblending of crowded fields. At the depth of these Surveys,
$\calD \simeq 400\Mpc$, the above magnitude correlations correspond
to $\theta \la 1^\prime$ and
could also have a significant effect on the angular
clustering and its interpretation on these small scales.

\subsection{An illustration}

\begin{figure}
\centering
\centerline{\epsfxsize=9truecm 
\epsfbox{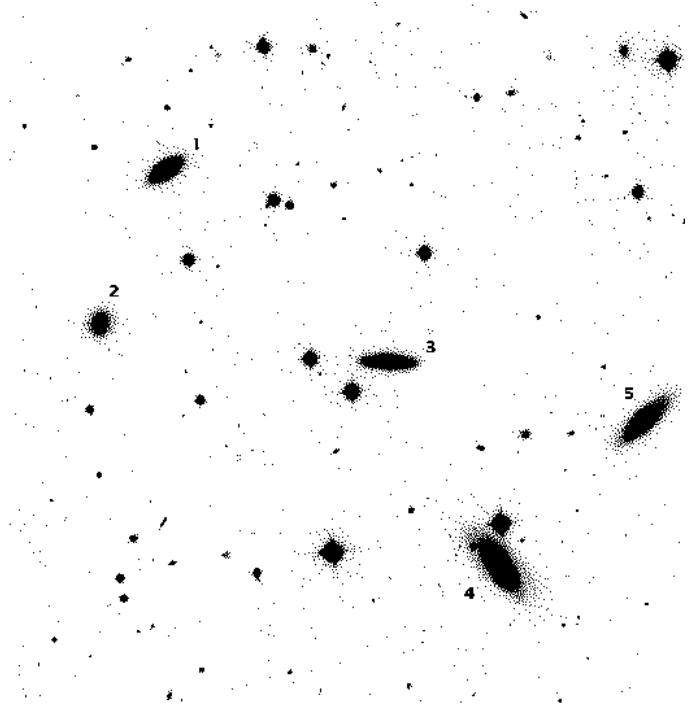}} 
\caption[junk]{A CCD image (in $B$) containing several Zwicky
galaxies (labeled with numbers) in a rich cluster.}
\label{int80a}
\end{figure}

\begin{figure}
\centering
\centerline{\epsfxsize=9truecm 
\epsfbox{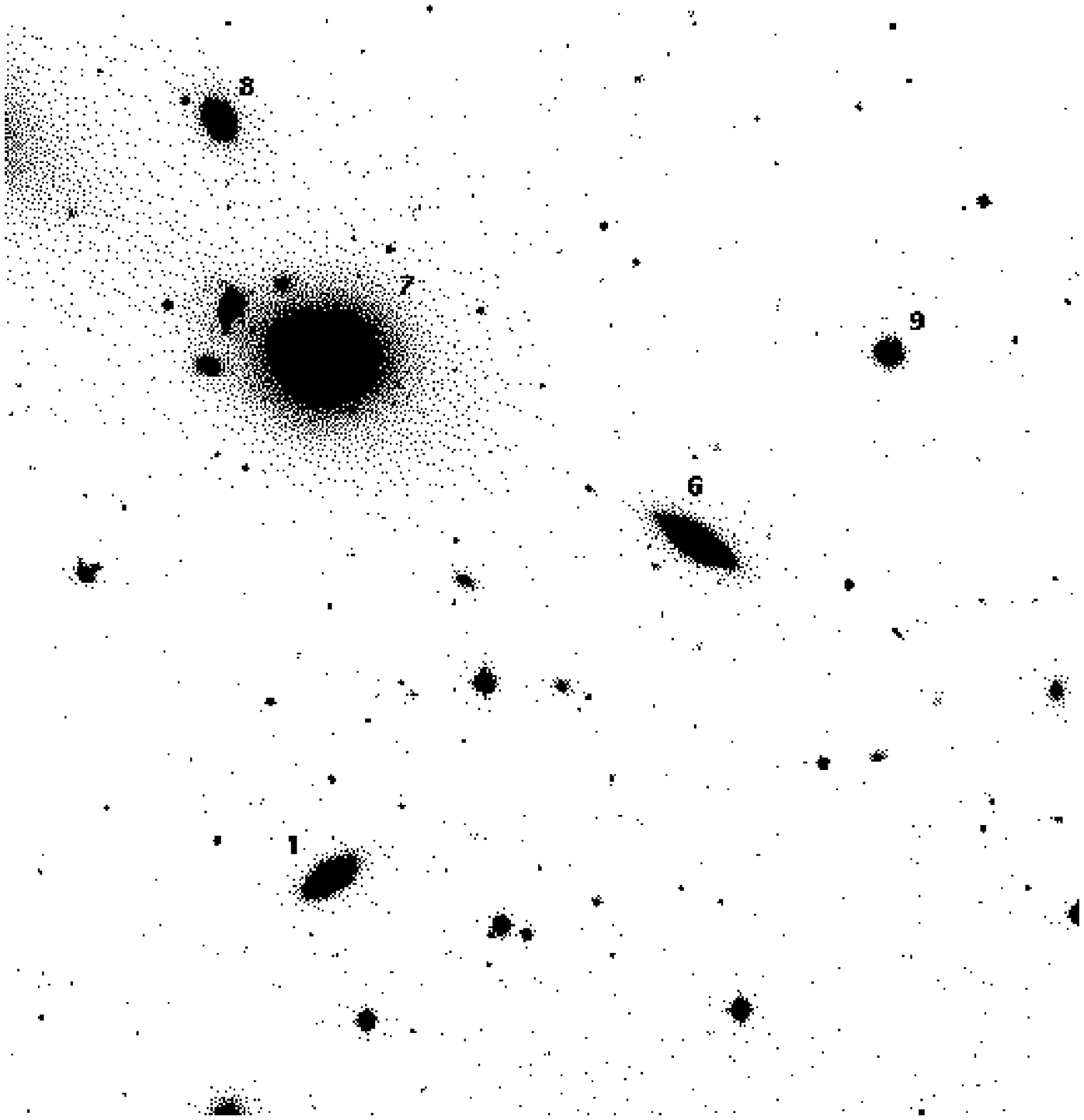}} 
\caption[junk]{A CCD image (in $B$) containing several Zwicky
galaxies (labeled with numbers) in a rich cluster.}
\label{int82a}
\end{figure}

An illustration of the scale error in the Zwicky system can be seem in
the galaxies shown in Figures \ref{int82a} and~\ref{int80a}.  Table
\ref{table1} gives the Zwicky and CCD magnitudes for each of the
Zwicky galaxies as labelled in the Figures.  As can be seen from the
table, the range of Zwicky galaxies in Figure \ref{int80a} is $\Delta
m_Z =0.3$ which is almost 6 times smaller than the CCD range $\Delta
B_{CCD} =1.7$. The range for the whole cluster is $\Delta m_Z =1.3$,
almost a factor of 3 smaller that the CCD range: $\Delta B_{CCD}
=3.5$. These illustrations are strongly suggestive of an observer-bias
effect, whereby some fainter galaxies are included in the same due to
their proximity to brighter objects, e.g. object~\#9 in Figure~\ref{int82a}.

\begin{table}
\begin{center}
\begin{tabular}{|c|c|c|}
Galaxy \# & Zwicky $m_Z$ & CCD $B$   \\ \hline 
1 & 15.0 & 15.0 \\ \hline 
2 & 15.0 & 16.1 \\ \hline 
3 & 14.8 & 15.1 \\ \hline 
4 & 14.7 & 14.4 \\ \hline 
5 & 14.8 & 14.9 \\ \hline 
6 & 14.9 & 14.8 \\ \hline 
7 & 14.0 & 13.2 \\ \hline  
8 & 15.2 & 14.9 \\ \hline  
9 & 15.3 & 16.7 \\ \hline 
\end{tabular}

\caption[junk]{Comparison of CCD and Zwicky magnitude for
galaxies, as labeled in Figures~\ref{int80a} and ~\ref{int82a}. }
\label{table1}
\end{center}

\end{table}

\section{The properties of nearby Field Galaxies}
\label{sec:proper}

We will now present some further properties of the galaxies in our
sample: colours, sizes and ellipticities, and discuss the
implications for the Luminosity function.
This will help us to understand the systematic effects present in the Zwicky
magnitude system. As mentioned in \S~1,
these local properties are interesting in the context of galaxy
evolution and star formation rates. 

There are still only few samples that are
both homogeneous and large enough to provide estimates of the
statistical properties of nearby samples, 
most of which have been selected from photographic
plates. Besides the redshift catalogues listed in Table 1,
one of the more extensive catalogues of bright galaxies
is contained in the {\it Second Reference Catalogue
of Bright Galaxies}
(de Vaucouleurs, de Vaucouleurs \& Corwin 1976, known as  RC2),
which gives magnitudes, colours, ellipticity and morphology for
well over a thousand galaxies to a limiting isophote of around $25.0 \magsq$. 
The problem with this catalogue is that it is a mere compilation of data
and was not intended  to be complete to any specified 
limiting magnitude, diameter, or redshift. Moreover it is based on 
photographic plates.

Our sample is homogeneous enough and extends over
a large enough area ( $\simeq 400$ square degrees) to provide a fair sample. 
Thus the new results based on the CCD magnitudes
should be a good local reference for the
fainter studies of galaxy evolution.

We will compare the properties of galaxies in the
 Zwicky ($m_Z<15.5$) sample with the corresponding properties of
galaxies in the INT fields selected using the CCD 
blue magnitude $B=B_{CCD}$. We choose two magnitude cuts: $B<16$
which includes most Zwicky galaxies and $19<B<20$, which
includes faint galaxies in the same fields.

\subsection{Sizes and Ellipticities}

\begin{figure}
\centering
\centerline{\epsfxsize=8.truecm 
\epsfbox{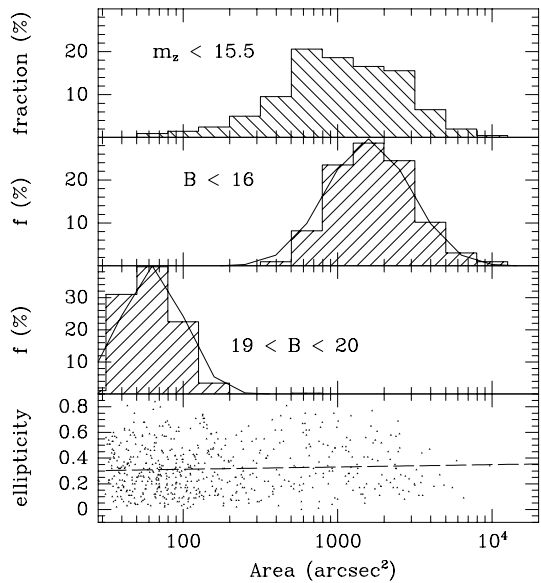}}
\caption[junk]{Frequency distribution (in per cent)
of galaxies as a
function of their area (in $\arcsq$) for galaxies selected in
different ways: (a) Zwicky sub-sample (top), 
(b) bright galaxies with CCD magnitudes $B<16$  (middle top) and
(c) faint galaxies with CCD magnitude $19<B<20$  (middle  bottom).
(d) area versus ellipticities for all CCD magnitudes
$14<B<20$ (bottom).}
\label{histarea}
\end{figure}

Figure \ref{histarea} shows an histogram of 
the frequency distribution of galaxies as a function
of its area (number of galaxy pixels above detection threshold in the CCD
image). Top panel shows Zwicky selected galaxies only. 
Middle and bottom panels include all the galaxies 
in the INT fields 
selected with CCD magnitudes of $B<16$ and $19<B<20$, respectively.

As can be seem in the Figure, Zwicky selected galaxies have a long
tail of small objects which is not present in the bright subsample
of $B<16$ CCD selected objects. 
This again illustrates the scale error (and selection biases) 
in the Zwicky system mentioned above (see also \S~6).

The distribution of sizes in logarithmic scale for both
 CCD sub-samples are well approximated by Gaussians (shown as
continuous lines in the figure). The mean size is about $60$ $\arcsq$
for $19<B<20$ and $1600$ $\arcsq$ for $B>16$, with a rms dispersion of
about $20\%$ and $27\%$ respectively
(recall that these areas correspond to a nominal threshold of
$26.1\magsq$, fainter thresholds could be needed to sample
the size of the lower surphace halos).

\begin{figure}
\centering
\centerline{\epsfxsize=7.truecm 
\epsfbox{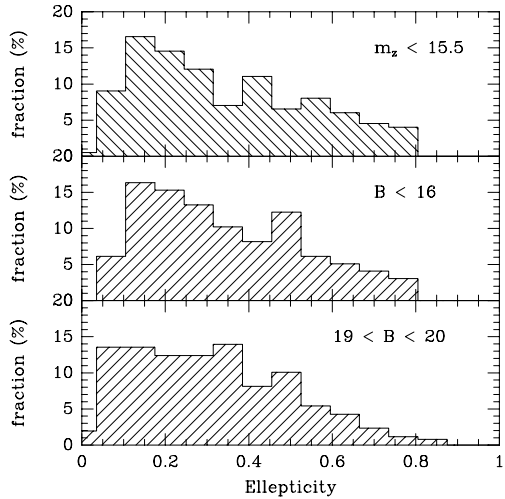}}
\caption[junk]{Frequency distribution (in per cent)
of galaxies as a
function of their ellipticity for galaxies selected in
different ways: (a) Zwicky sub-sample (top), 
(b) bright galaxies with CCD magnitudes $B<16$  (middle) and
(c) faint galaxies with CCD magnitude $19<B<20$  (bottom).}
\label{histell}
\end{figure}

Figure \ref{histell} shows the corresponding frequency distribution
for the ellipticities as measure in the galaxy shapes
(with a threshold of $26.1\magsq$). 
These Figures are in rough agreement with the results by  
Binney \& Vaucouleurs (1982) over 
the Second Reference Catalogue (RC2). 

The local distribution seems remarkably similar to the faint one, given
the large differences in sizes shown in Fig.\ref{histarea}. 
This is a good indication
that our isophote ($26.1\magsq$) is low enough, as otherwise we would
expect a large excess of round faint objects.
There seems to be  nevertheless a slight 
($\simeq 5\%$) excess of round faint objects. This is probably not 
due to the seeing (or pixel resolution) as most of the faint objects here
have more than 30 $\arcsq$ (or 100 pixels) 
of area (see Fig.\ref{histarea}).
The lack of correlation between ellipticities $\epsilon$ and areas ${\cal A}$
is  illustrated in the bottom panel of Figure \ref{histarea}. 
A least-square-fit to the points give $\epsilon \simeq 0.277 + 0.018
\log_{10} ({\cal A})$.


\subsection{Colours}

Both the morphology and a more detailed study of colours will be
presented elsewhere. Here we just show the colour frequencies
and discuss some of its implications. 

\begin{figure}
\centering
\centerline{\epsfxsize=7.truecm 
\epsfbox{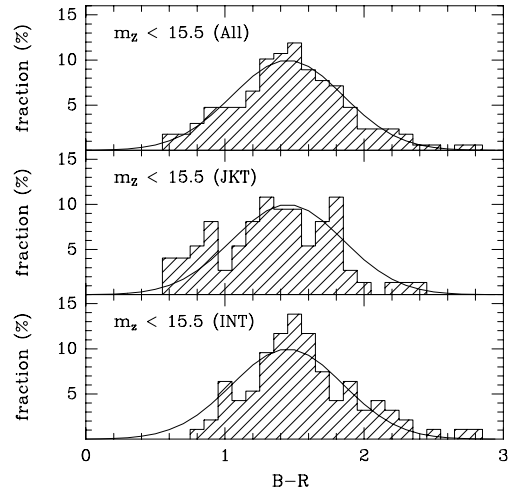}}
\caption[junk]{Frequency distribution (in per cent) of galaxies as a
function of CCD colour $B-R$ for Zwicky galaxies selected in different
ways: (a) All Zwicky in our sample (top), (b) Zwicky galaxies in our
JKT sub-sample (middle) (c) Zwicky galaxies in our INT sub-sample
(bottom). The same Gaussian distribution is plotted as a solid line in
each panel}
\label{histbr2}
\end{figure}

Figure \ref{histbr2} shows the frequency distribution of 
CCD colour ($B-R$) for all Zwicky galaxies (top) in our sample.
We then show the corresponding distributions 
for the same Zwicky galaxies separated into two sets,
corresponding to the JKT (middle) and INT (bottom) CCD
frames. As mentioned above, the INT field of view (10') was
larger than the JKT (6') and so we used the INT to target
groups and clusters of galaxies (which could take up several 
overlaping CCD frames), while the JKT was used
for more isolated galaxies. As can be seem in the Figure the
INT galaxies are significantly redder, $B-R \simeq 1.57$,
than the JKT, $B-R \simeq 1.36$. There are about $\simeq 100$
objects in each subset which roughly corresponds to a
random sampling of the total 600 Zwicky galaxies in our survey area
(the 600 targets were split in half between the two fields and then
more or less randomly selected during each night).
The mean colour of our 200 Zwicky galaxies is $B-R \simeq  1.47$.
Neither the mean colour or the frequency distribution change
significantly when we exclude all the galaxies that are in 
the same CCD frame (eg to avoid nearby ellipticals). This indicate that
selection effects are not important for this distribution.

Given the large area covered (several hundreds of square degree), these
results should be a good estimate of the overall mean colour of bright
local galaxies. How do these values compare with previous studies?
This is a difficult question because it requieres both accurate
colours and the accurate fraction of galaxies in different enviroments
(eg different morphological types).  Previous studies were limited to
inhomogeneous samples (eg RC2) or to
small numbers of galaxies.  For example Kennicutt (1992) presented
results for 8 early-type galaxies and 17 spirals to irregulars. Our
mean $B-R \simeq 1.47$ can be compared againts the synthetic $B-R$
colours presented by Fukugita et al. (1995). We have used Harris
filters which were designed to be very close to the Johnson--Cousins
(with $R$ Johnson) system. In these broadbands, Table 3 of Fukugita et
al. show $B_R=1.67$ for Ellipticals, $1.48$ for S0, $1.2-1.1$ for Spirals
and $0.6$ for Irregulars.  The range seems in rough agreement with
Figure \ref{histbr2}, although we find a significant number of objects
with $B-R>2$.  Notice that these objects are mostly in the INT sample,
that is in clusters or groups.  While most of the bluer objects with
$B-R<1$ are in the JKT fields, that is, they are more isolated (by
angular separations larger than $ 6'$, which corresponds to a mean
separation larger than $140h^{-1}\;{\rm Kpc}$.
Nevertheless, this is seems to be a small effect.

Note that Fukugita et al synthetic
colours seem to be slightly less red than the Kennicutt (1992)
observations (according to Table 2 in Fukugita et al., they are about
0.1 redder in B-V and this could be larger in B-R). Also notice that
we are using total magnitudes, while the synthetic colours are based
in small aperture spectra. Galaxies could have quite a different
colour distribution (or spectra) in their low-surface halos.

\begin{figure}
\centering
\centerline{\epsfxsize=7.truecm 
\epsfbox{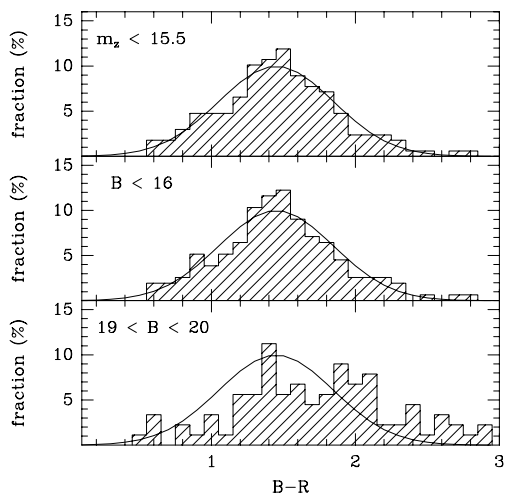}}
\caption[junk]{Frequency distribution (in per cent)
of galaxies as a 
function of CCD colour $B-R$ for galaxies selected in
different ways: (a) Zwicky sub-sample (top), 
(b) bright galaxies with CCD magnitudes $B<16$  (middle) and
(c) faint galaxies with CCD magnitude $19<B<20$  (bottom).}
\label{histbr}
\end{figure}

Figure \ref{histbr} shows the frequency distribution of 
CCD colour ($B-R$) for the Zwicky galaxies
(top panel) in comparison  with the same galaxies 
selected with CCD magnitudes: with $B<16$ (middle panel) and 
a sub-set of INT galaxies with $19<B<20$
(bottom panel), which corresponds to the points in Figure \ref{intrb}.
The continuous line in all cases
shows for reference a Gaussian
distribution with mean $B-R=1.45$ and {\it rms} deviation 
of $\sigma=0.4$, which roughly
matches the Zwicky frequencies. 

It is interesting to notice the peak of local galaxies
with $B-R=1.5$  and the spread and
shift to the red of the faint objects 
in the bottom panel. These magnitudes are not k-corrected,
which could well explain the relative redening of the faint
objects.

\subsection{The local CfA2 Luminosity Function (LF)}
\label{sec:LF}

Zwicky magnitudes have been used to estimate the LF in the
CfA2 redshift catalogue (Marzke et al. 1994). Marzke et al. used 
a Monte Carlo method to estimate how the LF parameters
would change if the dispersion $\sigma_M$ were $0.65\mag$ (closer to
what we find here than the nominal $0.3\mag$ they used). They found that
the {\it true} $M_*$ should be about $0.6\mag$ brighter, and that a
similar conclusion would be reached for a small scale error. Marzke et
al. used this estimate to conclude that a combination of incompleteness
and a small ($0.2\mag$) scale error would be sufficient to move the
CfA2-North values to those found from CfA2-South.

A detailed analysis of the implications of our new Zwicky magnitude
calibration on the 
LF would require the redshift information and this is left for future work.
It is nevertheless possible to use the mean relation that we
found to show how we expect the LF to change. 

In practice, when fitting the Schechter parameters in equation
\ref{schechter} to observational data, the value
of $\phi_*$ is correlated with the values of $M_*$ and $\alpha$.
In our case we do not use a fit to data, but 
just model how the LF changes with an homogeneous linear
shift in the magnitude scale. A zero point shift $Z_0$
 in the magnitude scale, as in
equation \ref{zero}, will just change the value of $M_*$:
\beq
M_*(corrected) = M_* + Z_0 . 
\eeq
As the LF measures the number density of galaxies per magnitude inverval,
a linear change in the magnitude scale will shift the amplitude
of the LF proportionally to the shift in the magnitude interval,
i.e. by  $\lambda$ in equation (\ref{scalerror}). Thus we have:
\beq
\phi_*(corrected) = \lambda \, \phi_*. 
\eeq

Thus the  scale and zero point
error in the Zwicky system give the following corrections for the CfA2
South results of Marzke et al. (1994) 

\beq
\phi_*(corrected) = \lambda \, \phi_* \simeq 0.0124 \pm 0.006 \, h^3\;{\rm Mpc}^{-3}
\eeq
\beq
M_*(corrected) = M_* + Z_0 \simeq  -19.3 \pm 0.3.
\eeq

\begin{figure}
\centerline{\epsfxsize=7.truecm \epsfbox{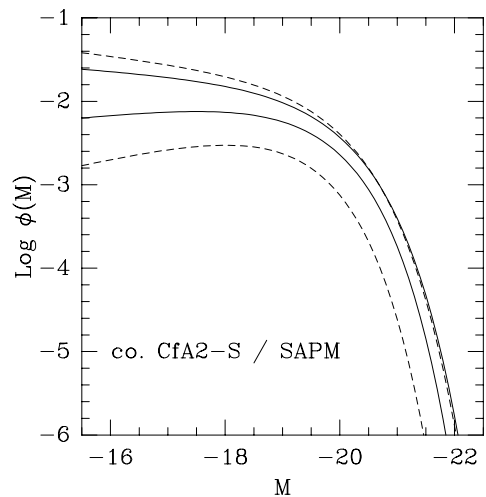}}
\centerline{\epsfxsize=7.truecm \epsfbox{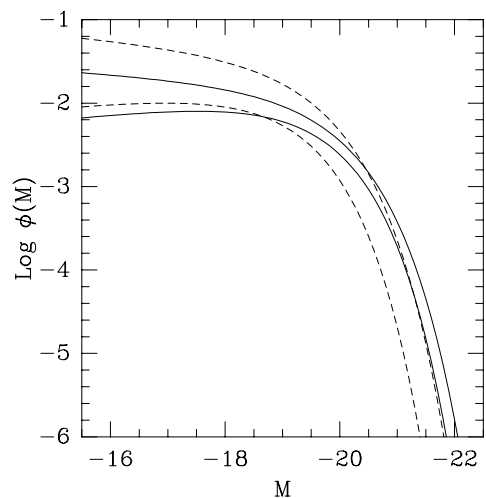}}
\caption{Luminosity function estimations.
The continuous lines enclose the 2-sigma region in
the SAPM estimation ($b_J$ band) whereas the dashed lines enclose the
2-sigma region in the CfA2 estimation ($B_{CCD}$) after correction for the model of
the Zwicky magnitude system and errors. The upper panel shows the
corrected LF for CfA2-South, and the lower for the whole CfA2 survey.}
\label{plotlf}
\end{figure}

If applied to the LF for the whole CfA2 survey:
\beq
\phi_*(corrected) = \lambda \, \phi_* \simeq 0.025 \pm 0.009 \, h^3\;{\rm Mpc}^{-3}
\eeq
\beq
M_*(corrected) = M_* + Z_0 \simeq  -19.1 \pm 0.2,
\eeq
where we have added the errors in quadrature. Thus the corrected values
are now closer the SAPM and LCRS (see Table \ref{tb:lfpars}). 
This can also be seem in Figure \ref{plotlf}, where we compare the 
corrected CfA2 luminosity function given above with the one corresponding
to the SAPM. Note that the later is in the APM $b_J$ band, so that there could be
some additional zero-point shifts between them (see \S 2).

\section{Conclusion}

In this paper we have presented CCD magnitudes for galaxies around
$204$  Zwicky galaxies which sample over 400 square degrees,
extending 6 hours in right ascension. This subsample is
drawn entirely from volume V of the Zwicky catalogue. 
We find evidence for a significant scale error as pointed
out by Bothun \& Schommer (1982) and Giovannelli \& Haynes (1984).
This is found by direct likelihood analysis, corrected for
Malmquist bias, and also by noticing the angular correlations between
the errors (Fig.\ref{nrzw}) or the long tail of small objects in the 
frequency distributions of sizes (Fig.\ref{histarea}).
The mean scale magnitude error is quite large: $\Delta
m_Z \simeq ~0.62 ~\Delta B_{CCD}$, i.e. an error of 
0.38 mag per magnitude, while the mean zero point is about
$-0.35$ (equation~\ref{zero}),
in agreement with Efstathiou {\it et al.} (1988).

An illustration of the scale error in the Zwicky system can be seem in
the galaxies shown in Figures \ref{int82a} and~\ref{int80a}. 
These illustrations are strongly suggestive of an observer-bias
effect, whereby some fainter galaxies are included in the sample due to
their proximity to brighter objects, e.g. object~\#9 in Figure~\ref{int82a}.
This bias can also be seem in statistical terms as a long tail
of small objects in the frequency distribution of Fig.\ref{histarea}.

Huchra (1976) found only a 0.08 mag per magnitude scale error in
a callibration of Zwicky galaxies 
with a photoelectric photometry of 181
sample of Markarian galaxies, which are preferentially spirals.
As spiral galaxies are bad tracers of clusters or groups, it is unlikely
that these Markarian galaxies include any of the fainter galaxies 
that contribute to the proximity observer-bias
mentioned above (which are mostly ellipticals). 
This effect would be hard to notice with photoelectric
photometry which samples only  one object at a time. 
Note that our analysis is restricted to a narrow range of magnitudes
$13.5< m_Z<15.5$, as compared to the wider range in Huchra (1976),
but this narrow range contains the majority of
the galaxies with $m_Z<15.5$ and therefore dominates all
the relevant statistical properties (such as the luminosity function).

Bothun \& Cornell (1990) have studied the calibration of the Zwicky
magnitude scale using a sample of 107 {\it spiral} galaxies. They
suggest that the errors in $m_Z$ are minimized if $m_Z$ is an
isophotal magnitude at $B=26.0\magsq$, although it is clear from their
Figure~2 that even within such a small sample of objects there is a
$5\mag$ range of isophotes which give isophotal magnitudes
corresponding to $m_Z$. This suggests that our isophotal detection
limit should be optimal for this comparison.

 Takamiya et
al. (1995), also with photoelectric photometry,
 find evidence for a large scale shift between Volume I and
Volumes II and V of the Zwicky catalogue. This shift appears to be of
order $0.5\mag\mag^{-1}$ over the range $14 < m_Z < 15.7$, which is
comparable to our findings for Volume V, although they find very
little effect for Volumes II and V. However, we note that Figure~4 of
Takamiya et al. shows $B-m_Z$ vs. $B$, rather than $m_Z$. A
similar representation of Figure~\ref{zwccdd} of this paper looks very
similar to Figure~4a of Takamiya et al., which is reasonable, given
that Figure~\ref{zwccdd} implies a strong compression of the $m_Z$
axis, which effectively hides the scale error.  Unfortunately, the
overlap between the 204 objects in our sample and the 155 objects in
the Takamiya et al. data is too small to draw any detailed comparison,
given that there are $\sim 600$ Zwicky galaxies within the region of
our survey.

We have also estimated how this scale error could change the CfA2 luminosity
function in \S\ref{sec:LF}. Figure \ref{plotlf} shows how
the corrected estimation is now closer to other local estimates, such as 
the SAPM.
Finally, in \S\ref{sec:proper} we give properties of the galaxies in our
sample: colours, sizes and ellipticities, providing one of the largest
samples of this kind.  
The local colour frequency distribution can be well approximated
by a Gaussian distribution with mean $B-R=1.45$ and rms deviation 
of $\sigma=0.40$. These colours compare well with synthetic $B-R$ colours 
presented Fukugita et al (1995). But there is a significant number of 
galaxies with  $B-R>2$ which are preferably found in clusters and groups,
while most of the  bluer objects with $B-R<1$ (spirals to irregulars) 
seem more isolated. These local properties are interesting in the 
context of galaxy evolution and star formation rates.
This will be studied in more detail in future work.

\medskip 
{\bf Acknowledgements} 
\medskip 

The Isaac Newton Telescope (INT) and Jacobus Kapteyn Telescope(JKT)
are operated on the island of La Palma by the Isaac Newton Group in
the Spanish Observatorio del Roque de los Muchachos of the Instituto
de Astrofisica de Canarias.  We thank Steve Maddox, Will Sutherland,
John Loveday, Cedric Lacey, Michael Vogeley, and Gary Wegner
for useful discussions. We also thank the referee for helpful
and constructive comments.  This work has been supported in part by
CSIC, DGICYT (Spain), projects PB93-0035 and PB96-0925, in part by
PPARC (UK), and by a
bilateral collaboration (Accion Integrada HB1996-0091) between CSIC
(Spain) and the British Council (UK).  Part of the data reduction and
analysis was carried out at the University of Oxford, using facilities
provided by the UK Starlink project, funded by PPARC.

\section{References}
\def\refe {\par \hangindent=.7cm \hangafter=1 \noindent}
\def\aj {ApJ,}
\def\na {Nature,}
\def\aa {A\&A,}
\def\prl {Phys.Rev.Lett.,}
\def\prd {Phys.Rev. D,}
\def\phr {Phys.Rep.}
\def\ajs{ApJS,}
\def\asj{A.J,}
\def\mn {MNRAS,}
\def\apl {Ap.J.(Let.),}

\refe Blair, M., \& Gilmore, G., 1982, PASP, 94, 7423
\refe Binney, J., \& de Vaucouleurs, G., 1981, MNRAS, 194, 679 
\refe Bothun, G.D., \& Cornell, M.E., 1990, AJ, 99, 1004
\refe Bothun, G.D., \& Schommer, R., 1982, ApJ, 255, L23 
\refe Da Costa, L.N., Geller, M.J., Pellegrini, P.S., Latham, D.W.,
Fairall, A.P., Marzke, R.O., Willmer, C.N.A., Huchra, J.P., Calderon,
J.H., Ramella, M., Kurtz, M.J., 1994, ApJ, 424, L1
\refe de Vaucouleurs, G., de Vaucouleurs, A., Corwin Jr., H.G., 1976,
Second Reference Catalogue of Bright Galaxies, University of Texas
Press, Austin.
\refe Efstathiou, G., Ellis, R.S., \& Peterson, B.J., 1988, \mn
{\bf 232}, 431
\refe Folkes, S., Ronen, S., Price, I., Lahav, O., Colless, M.M.,
Maddox, S.J., Deeley, K., Glazebrook, K., Bland-Hawthorn, J., Cannon,
R.D., Cole, S.M., Collins, C.A., Couch, W.J., Driver, S.P., Dalton,
G.B., Efstathiou, G., Ellis, R.S., Frenk, C.S., Kaiser, N., Lewis,
I.J., Lumsden, S.L., Peacock, J.A., Peterson, B.A., Sutherland, W.J.,
Taylor, K., 1999, MNRAS, 208, 459
\refe Fukugita, M., Shimasaku, K., Ichikawa, 1995, T., PASP, 107, 945
\refe Geller, M.J., Kurtz, M.J., Wegner, G., Thorstensen, J.R.,
Fabricant, D.G., Marzke, R.O., Huchra, J.P., Schild, R.E., Falco,
E.E., 1997, AJ, 114, 2205
\refe Giovanelli, R., \& Haynes, M.P., 1984, AJ, 89, 1
\refe Hamuy, M., \& Maza, J., 1989, AJ, 97, 720
\refe Huchra, J.P. 1976, \asj 81, 952
\refe Huchra, J.P., Davis, M., Latham, D. \& Tonry, J. 1983, ApJS, 52, 89
\refe Kennicutt, R.C., 1992, ApJS, 79, 255
\refe Landolt, A.U., 1992, AJ, 104, 340
\refe Lin, H., Kirshner, R.P., Shectman. S.A., Landy, S.D.,
Oemler, A., Tucker, D.L., Schechter, P.L., \aj {\bf 464}, 60
\refe Loveday, J., Peterson, B.A., Efstathiou, G., \& Maddox, S.J.,
1992, \aj {\bf 390}, 338
\refe Maddox, S.J., Sutherland, W.J., Efstathiou, G., \& Loveday, J.,
1990a, \mn {\bf 243}, 692
\refe Maddox, S.J., Sutherland, W.J., Efstathiou, G., \& Loveday, J.,
1990b, \mn {\bf 246}, 433
\refe Maddox, S.J., Sutherland, W.J., Efstathiou, G., \& Loveday, J.,
1991, {\it in ``The Early Observable Universe from Diffuse
Backgrounds''}, eds. Rocca-Volmerange, B., Deharveng, J.M., \&
Tr\^{a}n Thanh V\^{a}n, J.
\refe Marzke, R.O., Huchra, J.P., \& Geller, M.J., 1994, 
\aj {\bf 428}, 43
\refe Metcalfe, N., Fong, R., \& Shanks, T., 1995, MNRAS, 274, 769
\refe Peterson, B.A., Ellis, R.S., Efstathiou, G., Shanks, T., Bean,
A.J., Fong, R., \& Zen-Long, Z., 1986, MNRAS, 221, 233
\refe Schechter, P., 1976, \aj {\bf 203}, 297
\refe Szapudi, I., Gazta\~naga, E., 1998, MNRAS 300, 493
\refe Takamiya, M., Kron, R.G., \& Kron, G.E., 1995, AJ, 110, 1083
\aa 342, 15.
\refe Zucca, E., Zamorani, G., Vettolani, G., Cappi, A., Merighi, R.,
Mignoli, M., Stirpe, G.M., MacGillivray, H., Collins, C., Balkowski,
C., Cayatte, V., Maurogordato, S., Proust, D., Chincarini, G., Guzzo,
L., Maccagni, D., Scaramella, R., Blanchard, A., Ramella, M., 1997,
A\&A, 326, 477
\refe Zwicky, F.,  Herzog, E., Wild, P., Karpowicz, M., \& Kowal,
C.T., 1968, {\it Catalogue of Galaxies and Clusters of Galaxies}.

\end{document}